\documentstyle[preprint,aps,epsfig,floats]{revtex}
\begin{document}
\def\be{\begin{equation}}
\def\ee{\end{equation}}
\def\bc{\begin{center}}
\def\ec{\end{center}}
\def\bea{\begin{eqnarray}}
\def\eea{\end{eqnarray}}
\draft
\title{Weighted Scale-Free Networks with Stochastic Weight Assignments}
\author{Dafang Zheng$^{1,}$\thanks
{e-mail: zheng@physik.uni-halle.de}, Steffen Trimper$^{1}$,
Bo Zheng$^{1,2}$ and P. M. Hui$^{3}$}

\address{$^{1}$ Fachbereich Physik, Martin-Luther-Universit\"{a}t, D-06099 Halle, Germany}
\address{$^{2}$ Department of  Physics, Zhejiang University, Hangzhou 310027, P.R. China}
\address{$^{3}$ Department of Physics, The Chinese University of Hong Kong,\\
Shatin, New Territories, Hong Kong}

\maketitle \thispagestyle{empty}
\begin{abstract}

We propose and study a model of weighted scale-free networks
incorporating a stochastic scheme for weight assignments to the
links, taking into account both the popularity and fitness of a
node. As the network grows the weights of links are driven either
by the connectivity with probability $p$ or by the fitness with
probability $1-p$. Results of numerical simulations show that the
total weight associated with a selected node exhibits a power law
distribution with an exponent $\sigma$, the value of which depends
on the probability $p$. The exponent $\sigma$ decreases
continuously as $p$ increases. For $p=0$, the total weight
distribution displays the same scaling behavior as that of the
connectivity distribution with $\sigma= \gamma = 3$, where $\gamma$
is the exponent characterizing the connectivity distribution. An
analytical expression for the total weight is derived so as to
explain the features observed in the numerical results. Numerical
results are also presented for a generalized model with a
fitness-dependent link formation mechanism.
\end{abstract}


\narrowtext
%\vskip2pc]
%\begin{multicols}{2}

\section{Introduction}

Many complex systems, including social,
biological, physical, economic, and computer systems, can be studied
using network models in which the
nodes represent the constituents and links or edges
represent the interactions between
constituents \cite{albert1,dorogovtsev1}.
One of important measures of the topological structure
of a network is its connectivity distribution $P(k)$,
which is defined as the probability that a randomly selected node has
exactly $k$ edges. In traditional
random graphs \cite{erdos,bollobas} as well as
in the small-world networks
\cite{watts,barth,watts1}
the connectivity distribution shows exponential decay in the tail.
However, empirical studies on
many real networks showed that the
connectivity distribution exhibits a power law behavior
$P(k) \sim k^{-\gamma}$ for large $k$
\cite{albert1,dorogovtsev1}.
Networks with power-law connectivity distributions are
called {\it scale-free} (SF) networks.
Typical examples of SF networks include the Internet
\cite{faloutsos,caldarelli,pastor},
World-Wide-Web \cite{albert2,huberman,broder},
scientific citations \cite{redner}, cells \cite{jeong1,jeong2},
the web of actors \cite{albert3}, and the web of human sexual contacts
\cite{liljeros}. The first model of SF networks was proposed by Barab\'{a}si and Albert (BA) \cite{barabasi1}.
In BA networks two important ingredients are included in order
to obtain power law behavior in the connectivity
distributions, namely the networks are continuously {\it growing}
by adding in new nodes as time evolves, and the newly added nodes
are {\it preferentially attached} to the highly connected nodes.
The idea of incorporating preferential attachment in a growing network
has led to proposals of a considerable number of
models of SF networks \cite{krapivsky1,dorogovtsev2,albert4,krapivsky2,mossa,newman1,rozenfeld}
(see also  Refs. \cite{albert1,dorogovtsev1} and references therein).

In most growing network models, all the links are
considered equivalent.  However, many real systems
display different interaction strengths between nodes.
It has been shown that in systems such as the social
acquaintance network \cite{granovetter}, the web of
scientists with collaborations \cite{newman2},
and ecosystems \cite{berlow}, links between nodes may be different
in their influence and the so-called the weak links
play an important role in governing the network's functions.
Therefore, real systems are best described by weighted growing networks
with non-uniform strengths of the links.
Only recently, a class of models of weighted growing networks was
proposed by Yook, Jeong and Barab\'{a}si (YJB) \cite{yook}.
In the basic weighted scale-free (WSF) model of YJB,
both the topology and the weight are
driven by the connectivity according to the preferential
attachment rule as the network grows.
It was found that the total weight distribution follows
a power law $P(w) \sim w^{-\sigma}$, with an exponent
$\sigma$ different from the connectivity
exponent $\gamma$. It was also shown analytically
that the different scaling behavior in the
weight and connectivity distributions are results of
strong logarithmic corrections, and asymptotically (i.e., in the
long time limit) the
weighted and un-weighted models are identical \cite{yook}.

In real systems one would expect that a link's weight and/or the
growth rate in the number of links 
of a node depend not only on the ``popularity" of the
node represented by the connectivity, but also on some intrinsic
quality of the node. The intrinsic quality can be collectively
represented by a parameter referred to as the ``fitness"
\cite{bianconi,ergun}. Besides popularity, the competitiveness of
a node in a network may depend, taking for example a node being 
an individual in a certain community, on the personality,
survival skills, character, etc..  A newly added node may take
into account of these factors beside popularity in their decision
on making connections with existing nodes and on the importance of
each of the established links. Clearly, there is always a spectrum
of personality among the nodes and therefore a distribution in the
fitness.  While one may argue that factors determining the
popularity may overlap with those in fitness, it is not uncommon
that popularity is not the major factor on the importance of a
connection.  For example, we often hear that a popular person
actually has very few good friends, and an influential and
powerful figure in a network may often be someone very difficult
to work with. In the present work, we generalize the WSF model of
YJB to study the effects of fitness. In our model, the weights
assigned to the newly added links are determined stochastically
either by the connectivity with probability $p$ or by the fitness
of nodes with probability $1-p$. The scaling behavior of the total
weight distribution is found to be highly sensitive to the weight
assignment mechanism through the parameter $p$.

The plan of the paper is as follows. In Sec. \ref{sec:model}, we
present our model and simulation results. In Sec.
\ref{sec:solution}, we derive an analytical expression between the
total weight and the total connectivity of a node and provide a
theoretical explanation on the features observed in the numerical
results. Results on a generalized model with a fitness-dependent
link formation mechanism are presented in Sec.
\ref{sec:discussion}, together with a summary of results.

%************************************************************
\section{The Model and Numerical Results}
\label{sec:model}

The topological structure of our model
follows that of the BA model of SF networks \cite{barabasi1}.
A small number ($m_{0}$) of nodes are created initially.
At each time step, a new node $j$ with $m$ ($m \leq m_{0}$) links
is added to the network. These $m$ links will connect to
$m$ pre-existing nodes in the system according to the preferential
attachment rule that the probability $\Pi_{i}$
of an existing node $i$ being selected for
connection is proportional to the total
number of links $k_{i}$ that node $i$ carries, i.e.,
\begin{equation}
\label{Probk}
\Pi_{i}=\frac{k_{i}}{\sum_{l}k_{l}}.
\end{equation}
The procedure creates a network with $N=t+m_{0}$ nodes and $mt$ links
after $t$ time steps. Geometrically,
the network displays a connectivity distribution with a
power law decay in the tail with an exponent $\gamma=3$,
regardless of the value of $m$ \cite{barabasi1,barabasi2}.

A weighted growing network is constructed by assigning weights to
the links as the network grows.
To incorporate a fitness-dependent weight assignment
mechanism, a fitness parameter $\eta_{i}$ is assigned
to each node \cite{bianconi,ergun}.  The fitness $\eta_{i}$ is
chosen randomly from a distribution $\rho(\eta)$,
which is assumed to be a uniform distribution
in the interval $[0, 1]$ for simplicity.
With probability $p$, each newly established
link $j \leftrightarrow i$ is
assigned a weight $w_{ji}$ ($= w_{ij}$) given by

\begin{equation}
\label{Weik}
w_{ji}=\frac{k_{i}}{\sum_{\{i'\}}k_{i'}},
\end{equation}
where $\sum_{\{i'\}}$ is a sum over the $m$
nodes to which the new node $j$ is connected.
With probability $1-p$, $w_{ji}$ is determined by the fitness through

\begin{equation}
\label{Weif}
w_{ji}=\frac{\eta_{i}}{\sum_{\{i' \}}\eta_{i'}}.
\end{equation}
In Eqs. (\ref{Weik}) and (\ref{Weif}), $w_{ji}$ is normalized so
that the sum of the weights for the $m$ new links is unity, i.e.,
$\sum_{\{i'\}}w_{ji'} = 1$ \cite{yook}.
For $p=1$, our model reduces to the basic
WSF model of YJB in which the weights are
driven by the connectivity alone \cite{yook}. For $p=0$,
the weights are driven entirely by the fitness of the nodes.
For $0 < p <1$, the present model provides a
possible stochastic weight assignment scheme in which a newly
added node, e.g. representing some newcomer into a web, considers
either the popularity, or the fitness of its connected neighbor in determining
the influence of such a connection.

We performed extensive numerical simulations on the model.  In our
simulations, we studied networks up to $N=5 \times 10^{5}$ nodes
with $m = m_{0} = 5$.  For each value of $p$, results are obtained
by averaging over 10 independent runs. First we study the total
weight distribution $P(w)$, which is defined as the probability
that a randomly selected node has a total weight $w$. The total
weight of a node $i$ is given by the sum of the weights of all
links connected to it, i.e., $w_{i}=\sum_{j}w_{ij}$. Figure 1
shows that $P(w)$ behaves as a power law $P(w)\sim w^{-\sigma}$,
with an exponent $\sigma$ that decreases from the value of $3$ at
$p=0$ continuously as $p$ increases. For $p=1$, $\sigma=2.4$, a
result in agreement with that of the WSF model of YJB \cite{yook}.
For $p=0$, $\sigma=3$ ($=\gamma$) showing that $P(w)$ follows the
same scaling behavior as $P(k)$.  YJB found that the scaling
behavior of $P(w)$ depends strongly on $m$ \cite{yook} in their
model.  In the present model, we found that the $m$-dependence
persists for all $p > 0$. Only when $p=0$, $\sigma$ becomes
independent of $m$.

It is also interesting to study the dynamical behavior of the
total weight $w_{i}(\eta_{i},t)$ of some node $i$ with fitness
$\eta_{i}$. Fig. 2 shows that $w_{i}(\eta_{i},t)$ grows with a
power law behavior with time with an exponent $\delta$ that
depends on $p$.  For $p > 0$, $\delta > \beta$, where $\beta =
1/2$ is the exponent characterizing the dynamical behavior of the
connectivity $k_{i}(t)$ \cite{barabasi1}. For $p=0$,
$w_{i}(\eta_{i},t)$ shows the same scaling behavior as $k_{i}(t)$
with $\delta = \beta =1/2$. For $0 < p < 1$, $\delta$ depends on
the node's fitness $\eta_{i}$. Thus, the total weight actually
shows a multi-scaling dynamical behavior in the range $0 < p <1$
\cite{bianconi}.

The probability distribution $P(w_{ij})$ of the weights 
$w_{ij}$ of individual
links is also worth investigating. To suppress statistical
fluctuations, Fig. 3 shows the cumulative distribution,
$P(x>w_{ij})$, instead of $P(w_{ij})$, on a log-linear scale. For
$p = 0$, $P(x>w_{ij})$ decays exponentially in the tail. Recall
that $P(w)$ and $w_{i}(\eta_{i},t)$ show identical behavior as
$P(k)$ and $k_{i}(t)$ for $p=0$ respectively,  
and the latter two quantities are not sensitive to 
the weight assignment scheme.  Here $P(x>w_{ij})$
shows an exponentially decaying behavior, implying that the
weighted and un-weighted models are not totally identical even for
$p = 0$. For $p > 0$, the tail deviates from an exponential
decaying form and decays faster as $p$ increases. For $p = 1$, we
recover the results in the YJB model \cite{yook}.

%***********************************************************
\section{Analytical Solution}
\label{sec:solution}

To understand the different behavior between $w_{i}(\eta_{i},t)$
and $k_{i}(t)$ (as well as between $P(w)$ and $P(k)$) found in
numerical simulations, we derive an analytical expression for the
total weight $w_{i}(\eta_{i},t)$ of a node $i$ with fitness
$\eta_{i}$ at time $t$. Following YJB \cite{yook},
$w_{i}(\eta_{i},t)$ can be expressed as

\begin{equation}
\label{Twei}
w_{i}(\eta_{i},t)=1+\int _{t_{i}^{0}}^{t}\int _{m}^{\infty }
\int_{0}^{1}
\tilde{P}_{i}(m,t')
w_{ji}(\eta_{l},k_{l})\varrho(k_{l})\rho(\eta_{l})d\eta_{l}dk_{l}dt',
\end{equation}
where $\tilde{P}_{i}(m,t)$ is the probability that node $i$ is
selected for connection to a new node $j$ at time $t$ for given
$m$ and it is related to $\Pi_{i}$ in Eq.(\ref{Probk}) by a factor
of $m$. Here, $t_{i}^{0}$ is the time at which the node $i$ has
been added to the system. $w_{ji}(\eta_{l},k_{l})$ is the weight
assigned to the link between node $j$ and node $i$. $\varrho(k)$
and $\rho(\eta)$ are the probability distributions of $k$ and
$\eta$, respectively. According to the stochastic weight
assignment scheme modelled by Eqs. (\ref{Weik}) and
(\ref{Weif}), the weight $w_{ji}(\eta_{l},k_{l})$, on the average,
can be written as

\begin{equation}
\label{Avwei}
w_{ji}(\eta_{l},k_{l})=p\frac{k_{i}}{k_{i}+k_{l}}
+(1-p)\frac{\eta_{i}}{\eta_{i}+\eta_{l}},
\end{equation}
for the simple case of $m=2$.
Generalization to arbitrary
value of $m$ is straightforward.

From the connectedness of the SF model, $\tilde{P}_{i}(m,t)$,
$\varrho(k)$ and $k_{i}(t)$ are given by \cite{barabasi2,yook}

\begin{equation}
\label{Pt}
\tilde{P}_{i}(m,t)=m\Pi_{i}=\frac{k_{i}(t)}{2t},
\end{equation}
\begin{equation}
\label{Varr}
\varrho(k)=mk^{-2},
\end{equation}
\begin{equation}
\label{Ki}
k_{i}(t)=\frac{m}{\sqrt{t_{i}^{0}}}\sqrt{t}.
\end{equation}
Substituting Eqs.(\ref{Avwei}) - (\ref{Ki}) into Eq.(\ref{Twei})
and noticing that $\rho(\eta)$ is assumed to be a uniform
distribution in the interval $[0,1]$,
the integration in Eq. (\ref{Twei}) can be carried out to give

\begin{equation}
\label{Finalwei}
w_{i}(\eta_{i},t)\simeq
[p+2(1-p)\eta_{i}\ln{\frac{1+\eta_{i}}{\eta_{i}}}]k_{i}(t)
-\frac{1}{4}p[(\ln{\frac{4t}{t_{i}^{0}}})^{2}
-4\ln{2}\ln{\frac{t}{t_{i}^{0}}}]+C,
\end{equation}
where $C$ is an integration constant.
Eq.(\ref{Finalwei}) implies that the different scaling behavior in
$w_{i}(\eta_{i},t)$ and $k_{i}(t)$ as shown in the simulations are
results of the logarithmic correction term, which can be tuned by
the parameter $p$. For $p \rightarrow 0$, Eq. (\ref{Finalwei})
gives

\begin{equation}
\label{Finalweif}
w_{i}(\eta_{i},t)\sim2\eta_{i}\ln{\frac{1+\eta_{i}}{\eta_{i}}}k_{i}(t),
\end{equation}
leading to the same scaling behavior of $w_{i}(\eta_{i},t)$ and
$k_{i}(t)$, as observed in the simulation results. For $p=1$
corresponding to the WSF model of YJB \cite{yook}, the dynamical
behavior of $w_{i}(\eta_{i},t)$ deviates most from that of
$k_{i}(t)$. For arbitrary $m$, $w_{i}(\eta_{i},t)$ follows a
similar form with $m$ dependence coming into the second term on
the right hand side of Eq.(\ref{Finalwei}).
%************************************************************
\section{Discussion}
\label{sec:discussion}

Our model can be easily generalized to allow for
a fitness-dependent link formation mechanism \cite{bianconi,ergun}.
In the basic model with fitness \cite{bianconi},
the probability $\Pi_{i}$ that a new link is established with
an existing node $i$ is determined jointly by the node's connectivity
$k_{i}$ and fitness $\eta_{i}$ with

\begin{equation}
\label{Probke}
\Pi_{i}=\frac{\eta_{i}k_{i}}{\sum_{l}\eta_{l}k_{l}}.
\end{equation}
To study the effects of fitness, we study a generalization of our model
by replacing Eq.(\ref{Probk}) by Eq.(\ref{Probke}) for link 
formation, while keeping
Eqs. (\ref{Weik}) and (\ref{Weif}) for
weights assignments.
The connectivity distribution follows a
generalized power law \cite{bianconi} with an
inverse logarithmic correction of the form

\begin{equation}
\label{Connectivity}
P(k)\sim\frac{1}{\log{k}}k^{-\gamma'},
\end{equation}
with $\gamma' = 2.255$.  Fig. 4 shows the numerical results
for the total weight distribution
$P(w)$ for three different values of $p = 0$, $0.5$ and $1$.
It is found that $P(w)$ follows the same
generalized power law form as $P(k)$, but with a different
exponent $\sigma'$ that depends on $p$. For $p > 0$,
$\sigma' < \gamma'$.  Only for $p = 0$, $P(w)$
and $P(k)$ have the same exponent of
$\sigma' = 2.25 \sim \gamma'$.  For the cumulative
distribution $P(x>w_{ij})$ of weights of individual links,
the numerical results are similar to those shown in Fig. 3.

In summary, we proposed and studied
a model of weighted scale-free networks in which the weights assigned
to links as the network grows are stochastically determined
by the connectivity of nodes with probability $p$ and
by the fitness of nodes with probability $1-p$.  The model leads to
a power law probability distribution
for the total weight characterized by an exponent $\sigma$
that is highly
sensitive to the probability $p$.
If the weight is driven solely by the fitness,
i.e., $p = 0$, $P(w)$ follows the
same scaling behavior of $P(k)$ with the same exponent $\sigma = \gamma$.
Similar results were also found in a generalized model
with a fitness dependent link formation mechanism.
An expression relating
the total weight and the total connectivity of a node was derived
analytically.  The analytical result was used to explain the features
observed in the results of numerical simulations.  In closing, we note that
although the total weight distribution $P(w)$ and the connectivity
distribution $P(k)$ carry different exponents $\sigma$ and $\gamma$
for $p > 0$ in our model, $P(w)$ still follows a
power law, i.e., has the same functional form as $P(k)$.
The same feature was also found in the generalized model.
However, one would expect that in some complex real systems even the
functional forms of $P(w)$ and $P(k)$ may be different.
It remains a challenge to introduce simple and yet non-trivial models
that give one behavior for the geometrical connection among the
constituents and another behavior for the extend of connectivity
between constituents.

\acknowledgments This work was supported by a DFG grant TR 3000/3-3.  One of us (P.M.H.)
acknowledges the support from the Research Grants Council of the
Hong Kong SAR Government under grant number CUHK4241/01P.

%************************************************************

%************************************************************
\newpage \centerline{\bf FIGURE CAPTIONS}

\bigskip
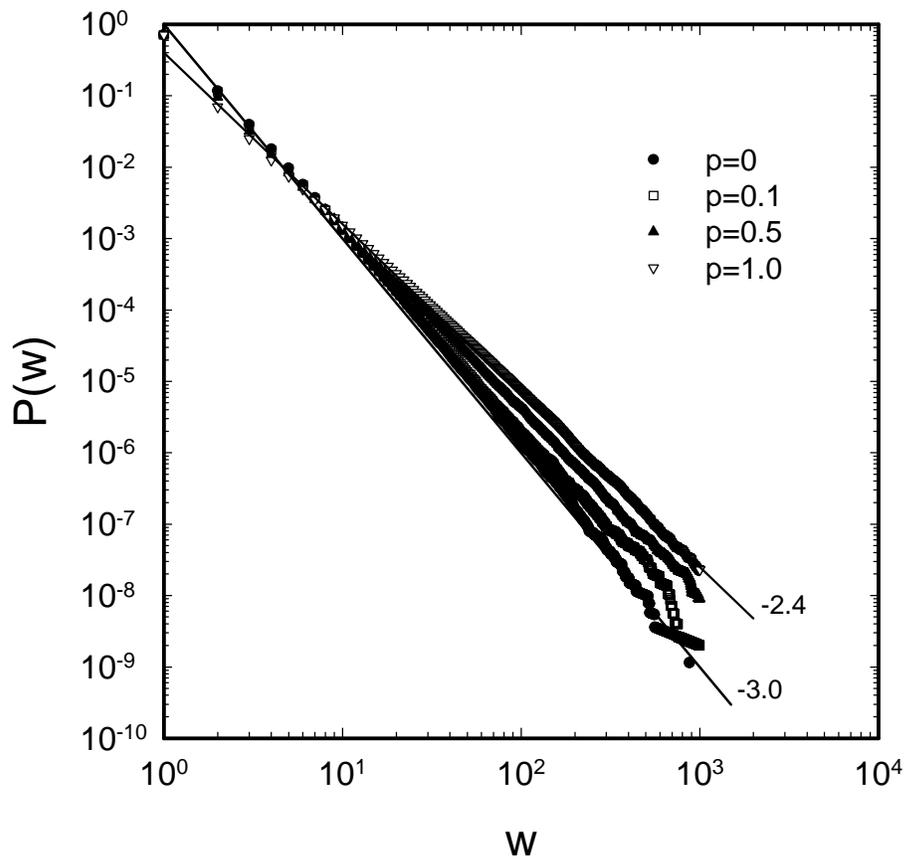
\noindent Figure 1: The weight distribution $P(w)$ as a function
of the total weight $w$ on a log-log scale for different values of
$p = 0, 0.1, 0.5, 1.0$. The two solid lines are guide to the eye
corresponding to the exponents $\sigma = 2.4$ and $3.0$, respectively.

\bigskip
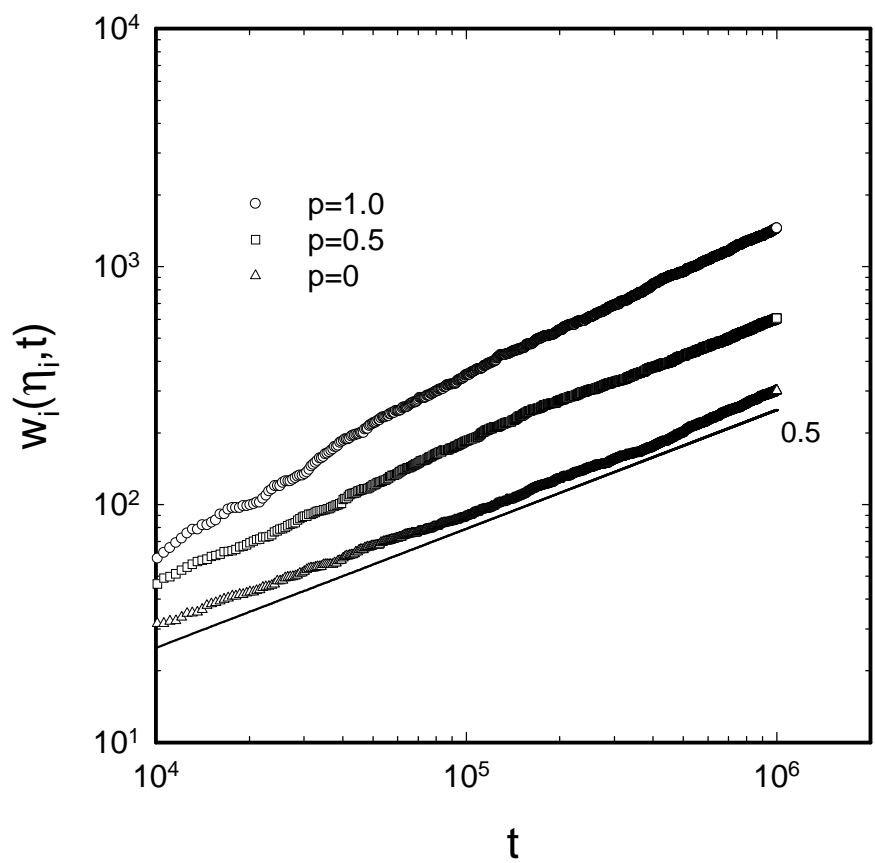
\noindent Figure 2:  The total weight $w_{i}(\eta_{i},t)$ of a
randomly selected node $i$ with fitness $\eta_{i}$ ($=0.75$) as a
function of time $t$ on a log-log scale for different values of $p
= 0, 0.5, 1.0$. The solid line is a guide to the eye corresponding
to an exponent $\sigma = 0.5$.

\bigskip
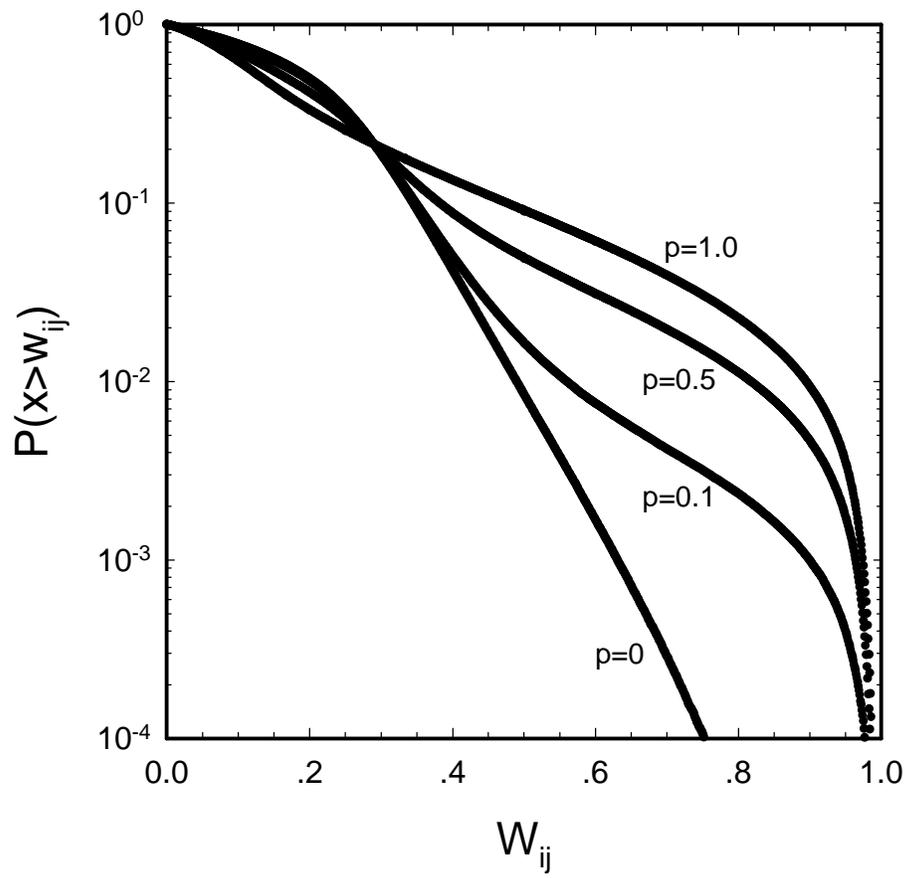
\noindent Figure 3: The cumulative distribution
$P(x>w_{ij})$ of the weights of individual links as
a function of $w_{ij}$ on a log-linear plot for different values of
$p = 0, 0.1, 0.5, 1.0$.

\bigskip
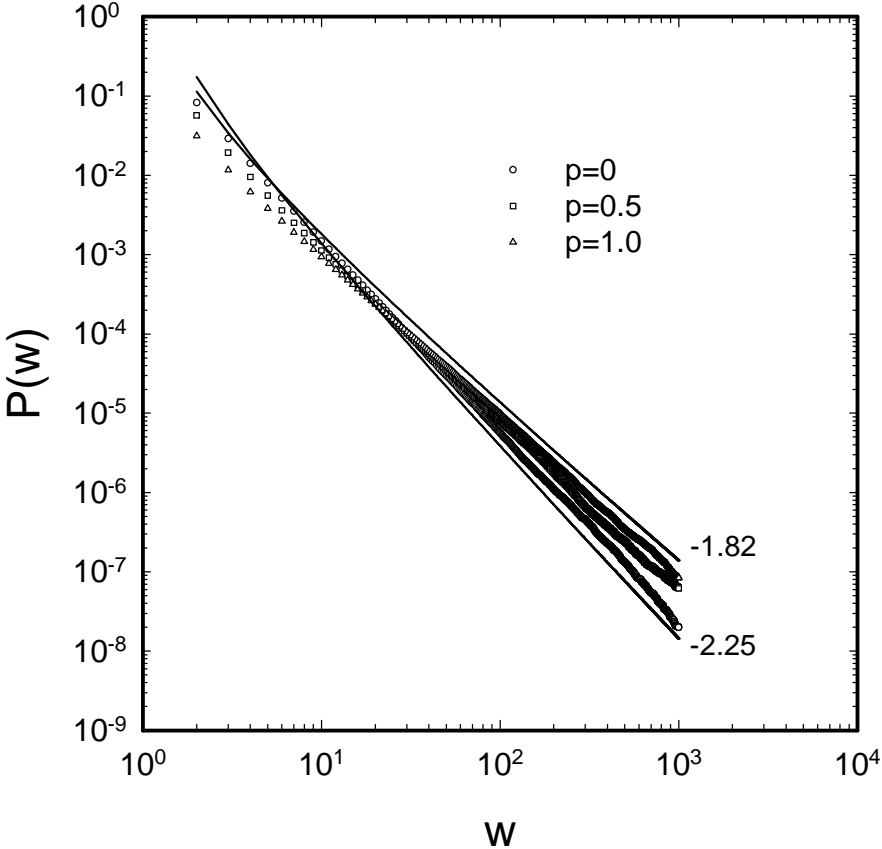
\noindent Figure 4: The weight distribution $P(w)$ as a function
of the total weight $w$ on a log-log scale for different values of
$p = 0, 0.5, 1.0$ in a model with fitness-dependent link formation
mechanism.
The two solid lines are plotted according to the form of
Eq.(\ref{Connectivity}), but 
with an exponent $\sigma'$ characterizing $P(w)$  
that takes on the values $1.82$ and $2.25$ respectively.

%\end{multicols}
\end{document}